\documentclass{PoS}

\title{Jet structure in integrated EPOS3-HQ approach}

\ShortTitle{Jet structure in EPOS3-HQ}

\author{\speaker{Iurii Karpenko}$^1$, Martin Rohrmoser$^2$, Pol Gossiaux$^1$, Joerg Aichelin$^1$, Klaus Werner$^1$\\
  $^1$ SUBATECH, IN2P3/CNRS, %
  Universit\'{e} de Nantes, IMT Atlantique, %
  4 rue Alfred Kastler, 44307 Nantes cedex 3, %
  France \\
  $^2$ Institute of Physics, Jan Kochanowski University, 25-406 Kielce, Poland\\
        E-mail: \email{yu.karpenko@gmail.com}}

\abstract{
We report on the first results for jet observables from a time-like parton cascade integrated with hydrodynamic evolution within the EPOS3-HQ framework. The hard (jet) partons are produced along with soft partons in the initial state EPOS approach. The soft partons, represented by strings, melt into a thermalized medium which is described by a 3 dimensional event-by-event viscous hydrodynamic approach. The jet partons then propagate in the hydrodynamically expanding medium. The total jet energy gets progressively ``degraded'' when the partons reaching a thermal energy scale are melted into the hydrodynamic medium via the source terms. The full evolution proceeds in a parallel mode, without separating the thermalized and jet parts. We demonstrate how the transverse expansion of the medium affects the jet structure, and how the medium itself is affected by the jet recoil.
}

\FullConference{International Conference on Hard and Electromagnetic Probes of High-Energy Nuclear Collisions\\
		30 September - 5 October 2018\\
		Aix-Les-Bains, Savoie, France}

\begin{document}

\section{Introduction}

Recent years brought big advancements in jet modeling in heavy ion collisions, in particular a greater understanding of in-medium modifications of the time-like jet parton shower. At the same time, the level of background medium modeling varies significantly from group to group. In recent studies, the QGP medium can be modeled as simple as a blob of constant temperature \cite{Zigic:2018smz} or as complex as an event-by-event 3+1 dimensional viscous hydrodynamically expanding system \cite{He:2018gxh}. However, it is becoming more understood now \cite{Cao:2018ews} that a precise modeling of the background medium evolution is an important component for quantitative description of jet structure in heavy ion collisions. In this report take a first look on the jet structure from a time-like parton cascade \cite{Rohrmoser:2018fkf} coupled with the EPOS3-HQ framework \cite{Werner:2013tya}. In particular we show how the transverse expansion of the medium affects the jet structure, and explore the back reaction of the jet to the medium.

\section{Model}
The study is carried out in the EPOS3-HQ framework \cite{Werner:2013tya}: both the initial state for the hydrodynamic phase and the initial hard partons are taken from \texttt{EPOS3} model (version 3.238). For the present study we have taken an event-averaged initial state, which leads to regular profiles of collective properties of the medium in space and time. This allows one to see the medium effects in a clearer way. Each initial hard parton leads to a development of a time-like parton cascade, due to collinear parton splitting caused by bremsstrahlung. The evolution of the parton cascade is performed with a Monte Carlo algorithm \cite{Rohrmoser:2018fkf} representing the Dokshitzer-Gribov-Lipatov-Altarelli-Parisi (DGLAP) equation with leading order $q\rightarrow qg$, $g\rightarrow gg$ and $g\rightarrow q\bar{q}$ splitting functions. The evolution of each parton cascade proceeds from an initial virtuality scale $Q_\uparrow$ given by the EPOS initial state down to a minimal virtuality scale of $Q_\downarrow=0.6$~GeV.

The DGLAP evolution proceeds in momentum space. In order to couple the parton cascade to the medium one has to make assumptions about the spacetime evolution of the former. Therefore, we assume that in the rest frame a parton has a mean life time (or the time before its next splitting occurs) $\Delta t=E/Q^2$. In the scenario with no collisional energy loss the partons propagate on straight trajectories defined by their momenta between the splittings.

We use vacuum splitting functions in the algorithm. The medium effects are added \`a-la YaJEM \cite{Renk:2008pp}. This means that the extra medium induced radiation results in a constant increase of virtuality squared for the off-mass-shell partons $\frac{dQ^2}{dt}=\hat{q}_R$,
where $t$ is time in the local fluid (medium) rest frame, and for the corresponding transport coefficient $\hat{q}_R$ we use the parametrization $\hat{q}_R=210\cdot T^3/(1+53\cdot T)$.
In the present study the effects of collisional energy loss are simulated exclusively in the form of longitudinal drag force $\frac{d \vec p}{dt} = -A\cdot \vec p/|\vec p|$,
where $\vec p$ is parton 3-momentum. The value of the longitudinal drag coefficient we choose as $\hat{q}_R/A=(0.09+0.715\cdot T/0.16)$~[GeV]. Such a choice is based on the Einstein-Smoluchowski relation between the transverse momentum transfer $\hat{q}_C$ and the drag force $A$, and assuming that $\hat{q}_C=\hat{q}_R$.

The expansion of the medium is described by a 3 dimensional relativistic viscous hydrodynamic approach using \texttt{vHLLE} code \cite{Karpenko:2013wva}. The equation of state (EoS) is an essential component of hydrodynamic modeling, and is also used to extract the temperature field. For this study we use an EoS table from \cite{Laine:2006cp}, which is compatible with the recent and widely used \texttt{s95p-v1.2} EoS \cite{Huovinen:2009yb} which is based on the UrQMD resonance list in the hadronic phase.

As the hydrodynamic evolution proceeds in timesteps, it is natural to use the same timesteps to chop the evolution of parton cascades. During each timestep the local temperature and collective flow velocity at the position of a parton are obtained by interpolation and a Lorentz boost of parton energy-momentum to the local rest frame is performed. Then the equations for the energy loss are solved for each parton in the corresponding local rest frame. The updated energy-momentum vector is then boosted back to the computational frame. The choice of having a splitting within a timestep is made with Monte Carlo sampling based on a current lifetime of the parton.

Partons loose their energy and momentum while propagating in the medium, therefore some of them may be slowed down to thermal energies. As the energy of a parton in the rest frame of the corresponding fluid element becomes smaller than the local temperature, the parton is ``melted'': its energy and momentum are distributed to nearby fluid cells and the parton is removed from the cascade.

Most importantly, the energy and momentum loss (or gain) of a parton is added to (or subtracted from) the fluid cells around the position of the parton, via source terms in the hydrodynamic equations. Henceforth we refer to this process as  `absorption'.

\section{Results}

\begin{figure}
 \includegraphics[width=0.52\textwidth]{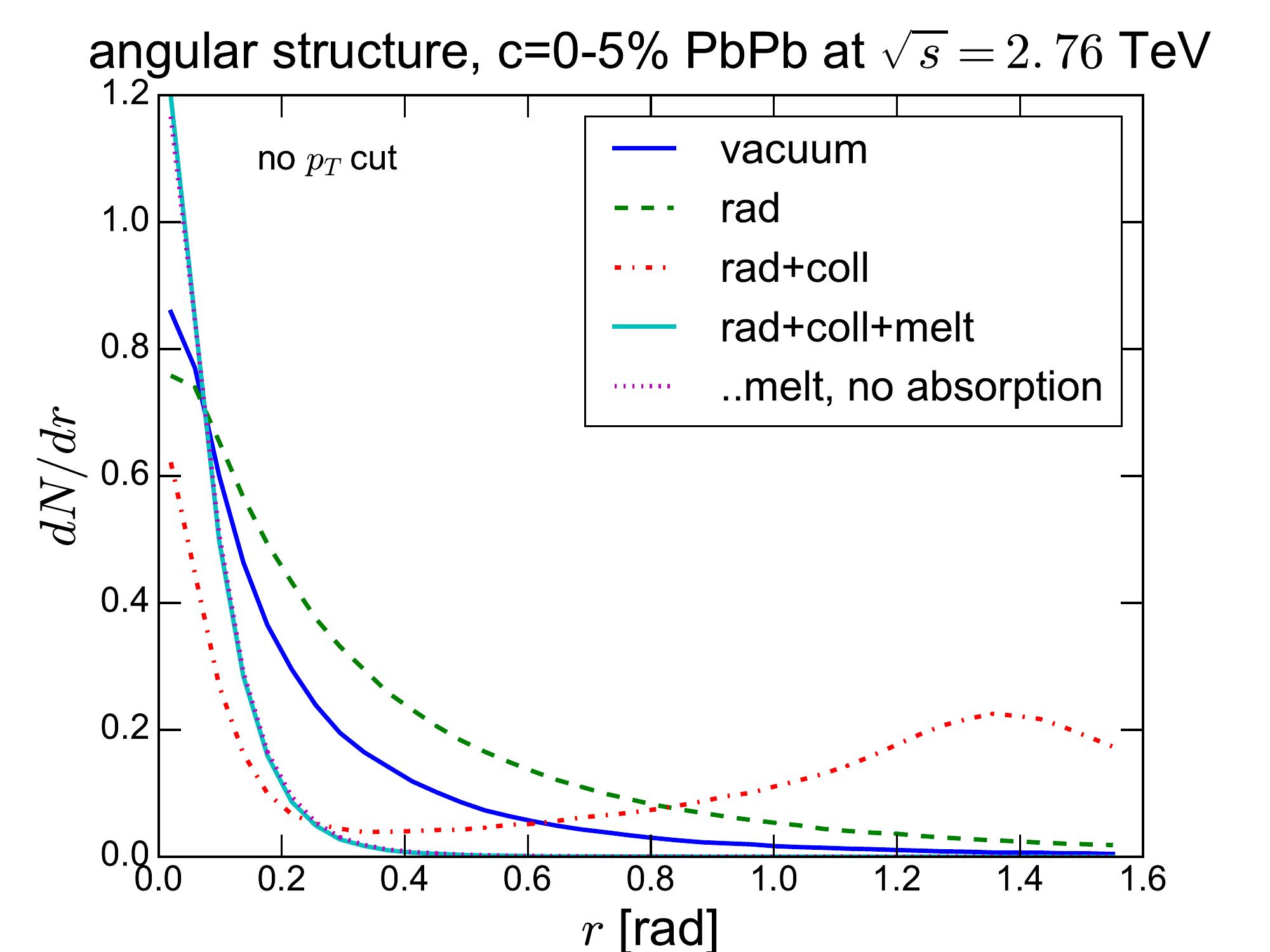}
 \includegraphics[width=0.52\textwidth]{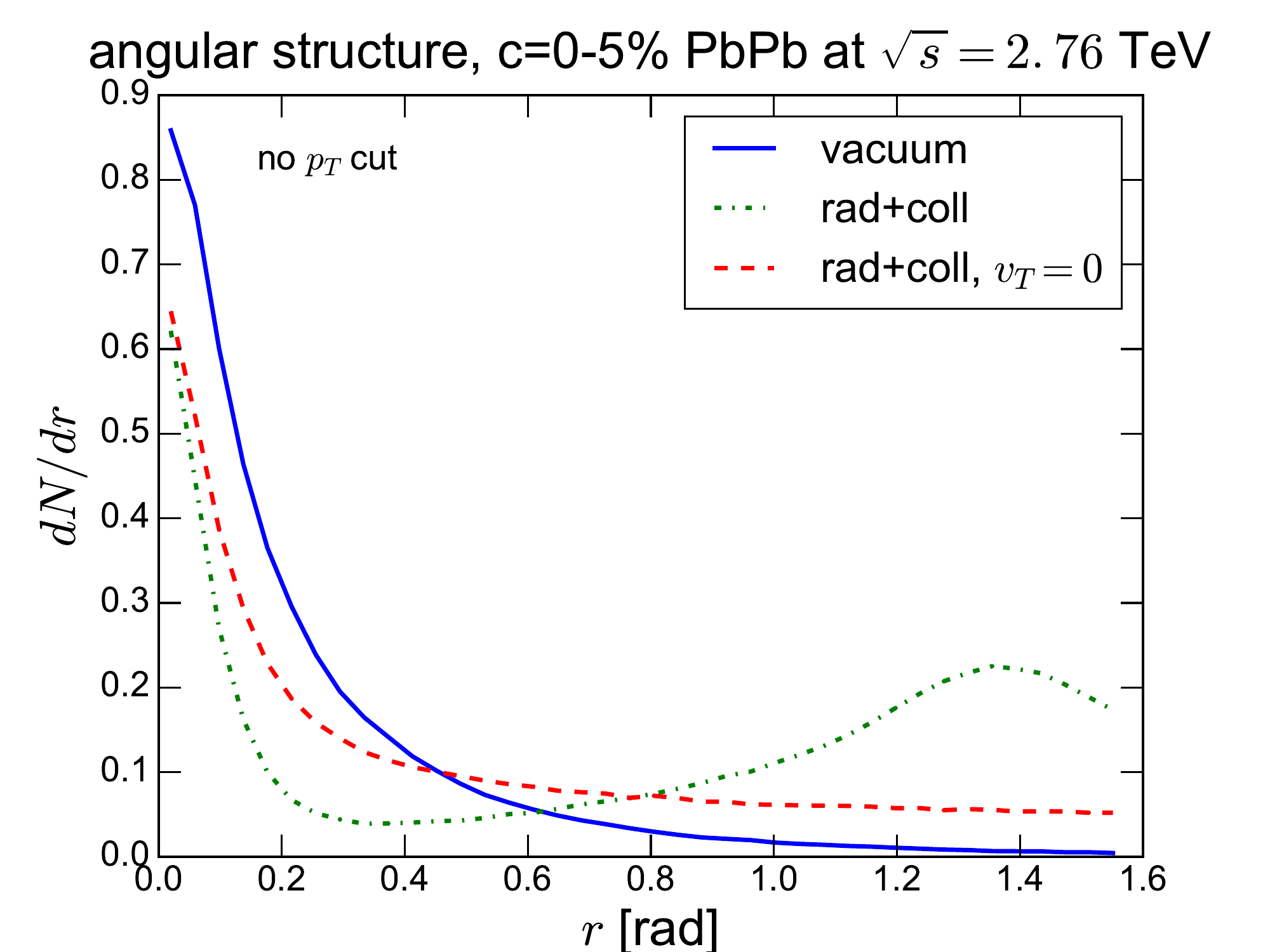}
 \caption{Angular distributions of jet partons within a jet cone. $r$ denotes an angle between the vector of a parton momentum and the direction of the initial jet parton. The curves represent different combinations of energy loss mechanisms (radiative, collisional and parton melting). See the text for more details.}\label{fig-jetang}
\end{figure}

Radiative and collisional energy loss mechanisms have somewhat different effects on the angular jet structure. This is shown on the left panel of Fig.~\ref{fig-jetang}. Radiative energy loss induces more gluon radiations by increasing the parton's virtuality between the splittings, thereby broadening the distribution of jet partons over the angle $r$ with respect to the direction of the jet axis. Inclusion of the longitudinal drag force shrinks the angular distribution at small angles $r$  but creates a broad structure at large $r$ peaking around $\pi/2$. We argue that such structure is induced by the radial flow of the expanding medium. This can be clearly seen on the right panel of Fig.~\ref{fig-jetang}, where we compare the calculations with radiative energy loss and longitudinal drag force when the transverse flow of the expanding medium is included (dash-dot curve) and when it is not (dashed curve). The 3 dimensional medium expansion is exactly the same in both cases - which implies the same radiative energy loss, whereas in the $v_T=0$ case a zero transverse velocity is assumed for the local rest frame boost procedure. This way the jet partons ``do not feel'' the transverse flow in the $v_T=0$ case. As one can see from the plot, the broad peak at $\pi/2$ is completely absent in the $v_T=0$ case.

One can also observe that in the full calculation (`rad+coll+melt' case on the left panel of Fig.~\ref{fig-jetang}) the parton melting procedure completely eliminates the large-$r$ structure. This happens because the deflected partons lose most of their momentum with respect to the fluid around and melt according to the criterion above. The energy-momentum loss perturbs the motion of the fluid (see below), and may show up as a corresponding enhancement in thermally produced hadrons.

\begin{figure}
 \includegraphics[width=0.52\textwidth]{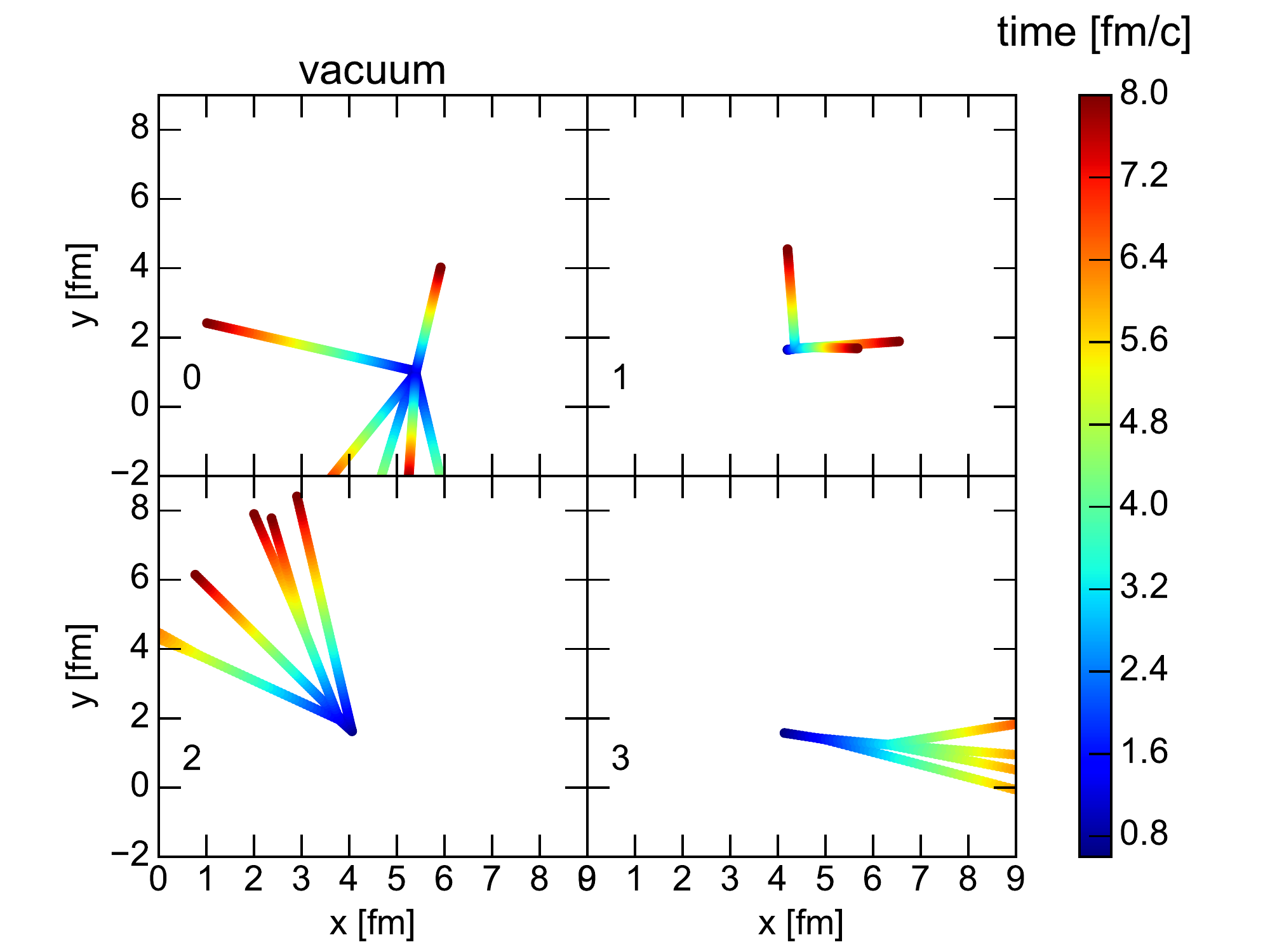}
 \includegraphics[width=0.52\textwidth]{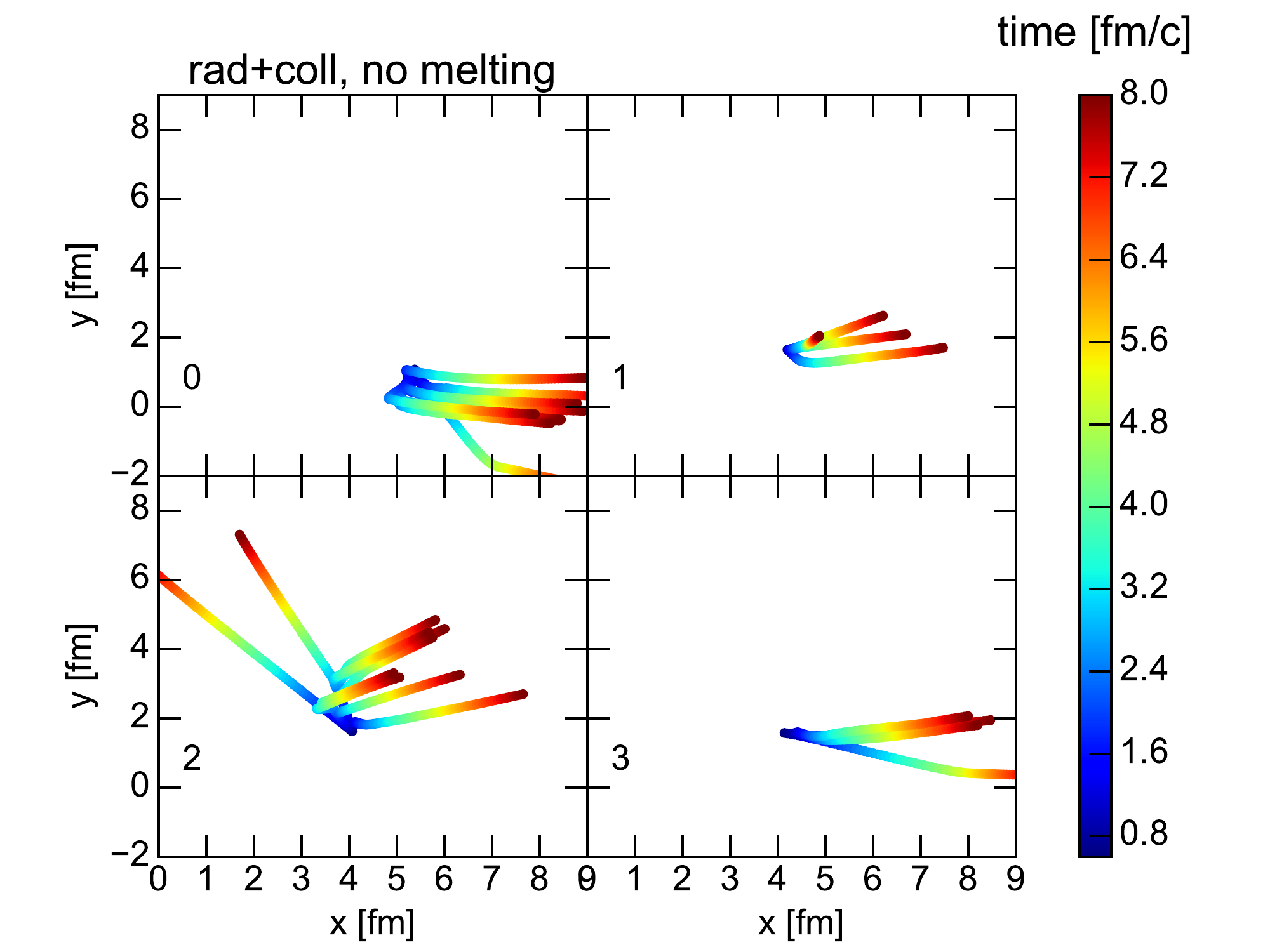}
 \vspace*{-20pt}
 \caption{Space-time trajectories of jet partons from 4 selected jets from a central 2.76 TeV PbPb event. The left panel shows the evolution without medium effects, whereas the right panel shows the evolution with medium effects. Color of the lines denotes time in the center-of-mass frame.}\label{fig-jetevo}
\end{figure}

The broad peak around $\pi/2$ looks less surprising if one follows the trajectories of individual jet partons in the medium. In Fig.\ \ref{fig-jetevo} we show the space-time evolution of 4 selected jets originating from partons with a relatively large initial virtuality $Q_\uparrow>10$~GeV in a central 2.76 TeV PbPb event, with (left panel) and without (right panel) medium effects. The parton trajectories in the `vacuum' case are obviously straight lines originating at each next splitting, whereas in the 'medium' case the trajectories are swept by the collective motion of the medium, which in this particular case moves in the +x direction. This happens because the longitudinal momentum drag is applied with respect to the rest frames of the fluid elements.

\begin{figure}
 \includegraphics[width=0.52\textwidth]{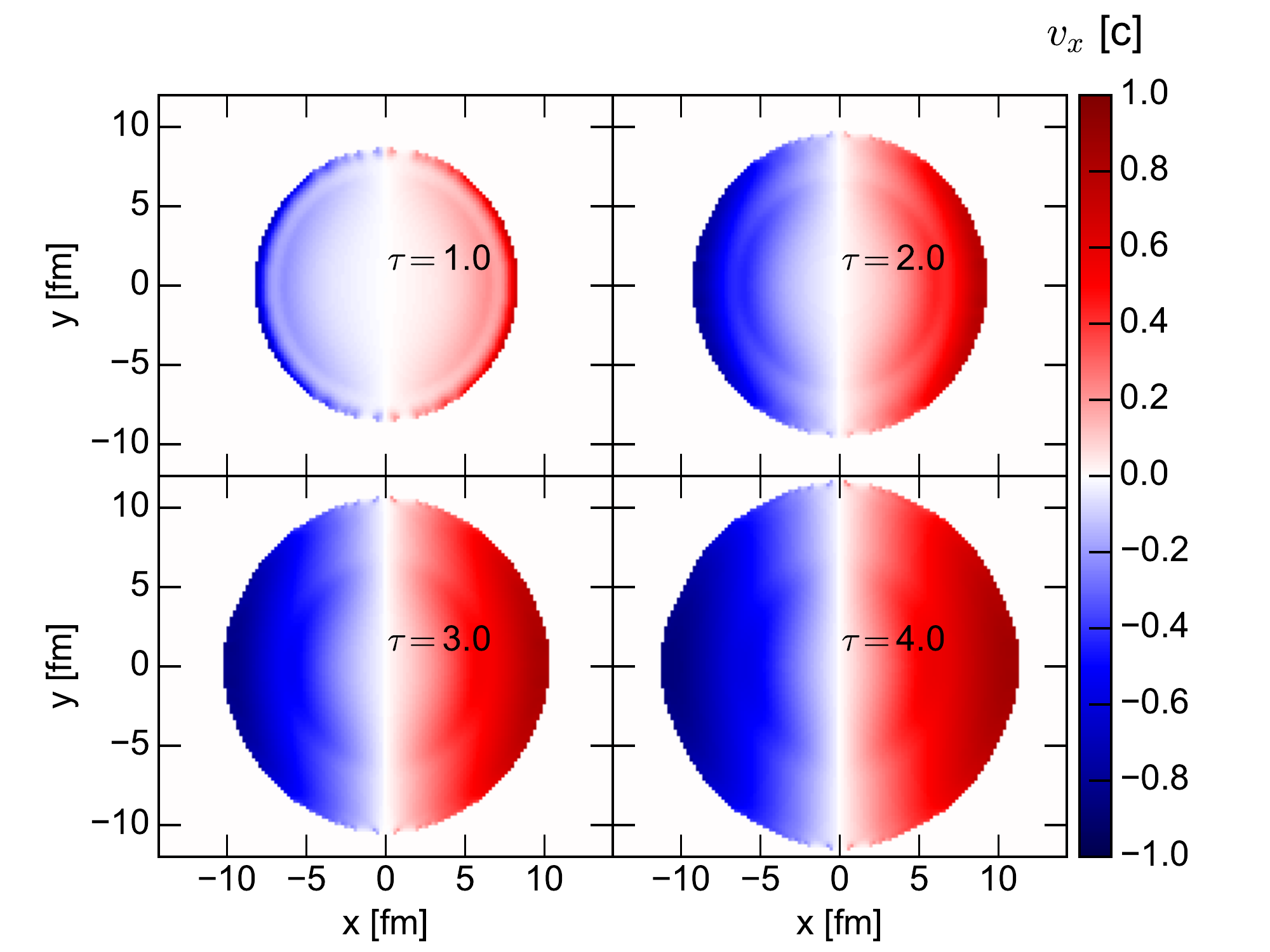}
 \includegraphics[width=0.52\textwidth]{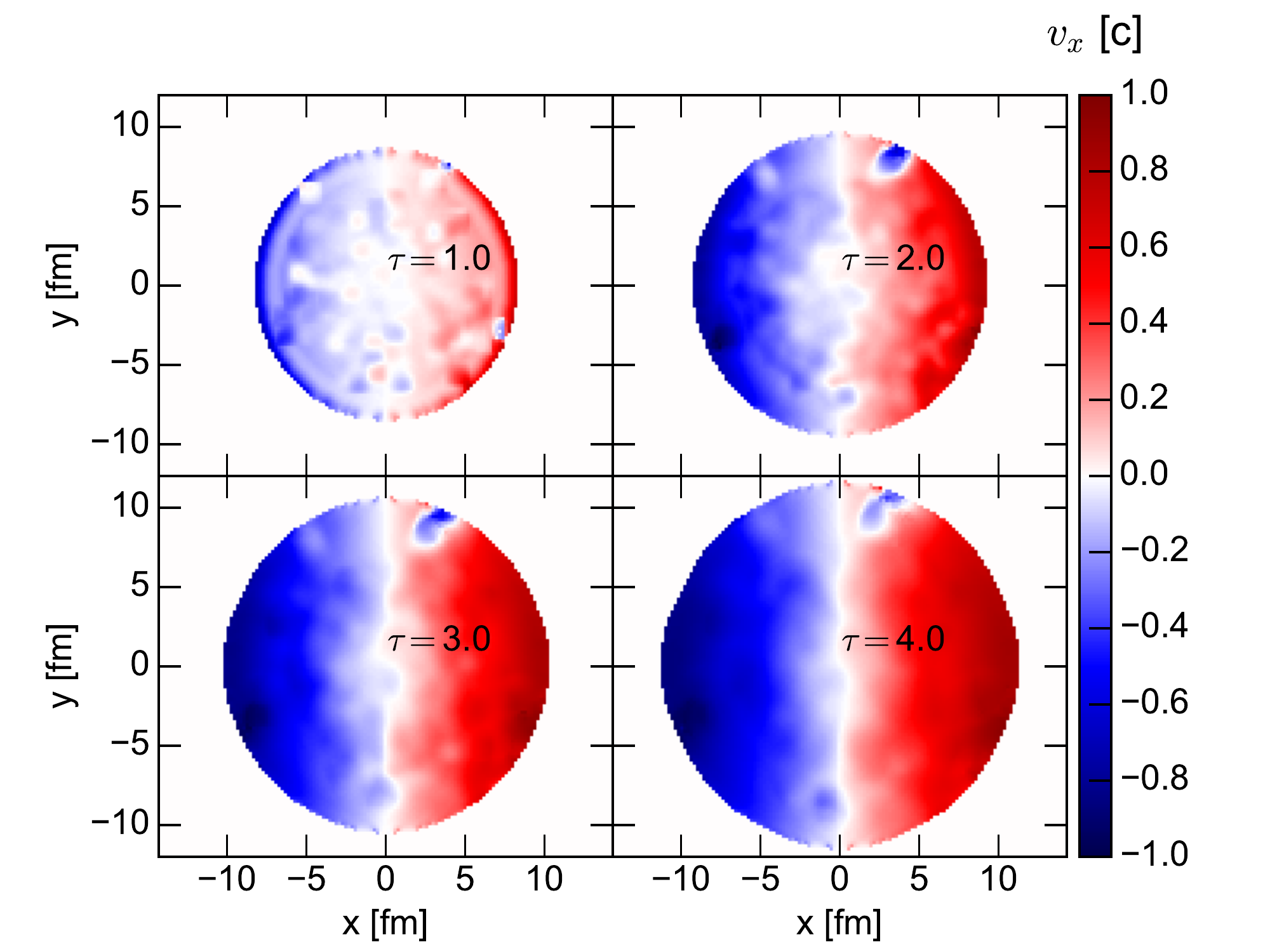}
 \vspace*{-20pt}
 \caption{X component of the collective flow velocity field in the transverse coordinate space, at different moments of time. Left panel: undisturbed hydrodynamic expansion, right panel: hydrodynamic expansion with absorption of the lost energy/momentum of partons. The hydrodynamic calculation is performed for average initial state corresponding to 0-5\% central PbPb collisions at $\sqrt{s_{\rm NN}}=2.76$~TeV.}\label{fig-vxevo}
\end{figure}

The energy and momentum loss by the jet partons causes perturbations in the hydrodynamic expansion of the medium. They can be seen most clearly in case of a hydrodynamic evolution starting from an averaged (therefore smooth and regular) initial state. In Fig.~\ref{fig-vxevo} one can see the development of $x$ component of velocity in the unperturbed (left panel) and perturbed (right panel) hydrodynamic evolution. The perturbed case corresponds to the calculation where the fluid 'absorbs' all the energy and momentum lost by partons via source terms in the hydrodynamic equations, thereby conserving the total energy of the hydrodynamic + jet system.

As one can see, the perturbations of the velocity field caused by the jet energy loss are largest at early times, whereas the late-time evolution at $\tau>4$~fm is almost regular and smooth. This happens because the medium-induced radiative energy loss coefficient and longitudinal drag force strongly increase with the temperature of the medium. At the same time, hydrodynamic expansion constantly works to dilute the created inhomogeneities.

Finally a few caveats must be noted. A hadronization algorithm was not used in this study, therefore all the distributions shown here are on the parton level. Consequently, a jet finding algorithm was not employed either. The work to include hadronization and jet finding algorithms is in progress.

Nevertheless we stress that the radial flow affects the jet structure at all $r$, therefore the radial flow of the medium must be included in a proper quantitative treatment of jets.

{\bf Acknowledgments}. The work is supported by Region Pays de la Loire (France) under contract no.~2015-08473. M.R.\ acknowledges support by the Polish National ScienceCentre (NCN) Grant 2015/19/B/ST2/00937

\end{document}